\documentclass[fleqn]{article}

\usepackage{axodraw}
\usepackage{subfigure}
\usepackage{graphicx}

\usepackage{latexsym}

\usepackage{amsmath, amssymb, graphics}

\newcommand{\mathsym}[1]{{}}

\begin{document}

\begin{titlepage}
\begin{flushright}
{  26 July 2007}
\end{flushright}

\vspace{15mm}

\begin{center}
{\LARGE {\bf Neutrino production states in oscillation phenomena -
are they pure or mixed?} }
\vspace{15mm} \\

 { \large {\bf Micha{\l} Ochman\footnote{mochman@us.edu.pl},  \
 Robert
 Szafron\footnote{rszafron@us.edu.pl}, \  {\normalsize and}  \
Marek  Zra{\l}ek}\footnote{zralek@us.edu.pl}} \\
\vspace{10mm}
{\sl Department of Field Theory and Particle Physics, \\ Institute of Physics,} \\
{\sl University of Silesia, Uniwersytecka 4, 40-007 Katowice, Poland}\\
{\sl Tel: 048 32 2583653, Fax: 048 32 2588431}

\end{center}

\setcounter{equation}{0} \vspace{15 mm}

\begin{abstract}
\vspace{5mm} \hspace{0.8cm}

General quantum mechanical states of neutrinos produced by
mechanisms outside the Standard Model are discussed. The neutrino state is described
by the Maki-Nakagawa-Sakata-Pontecorvo unitary mixing matrix only in the
case of relativistic neutrinos and Standard Model left-handed
charge current interaction. The problem of Wigner spin rotation caused by Lorentz transformation from the rest production frame to the laboratory frame is
considered. Moreover, the mixture of the neutrino states as a function of
their energy and parameters from the extension of the Standard Model are
investigated. Two sources of mixture, the appearance of
subdominant helicity states, and mass mixing with several
different mixing matrices are studied.

\end{abstract}

\vspace{3mm}

 \\
 \\  \ \

 \vspace{2mm} \vfill \vspace{5mm}
\end{titlepage}

\section{Introduction}
\label{INTR} The neutrino oscillation phenomenon is well established.
In several different experiments, the transition between different
neutrino flavours is observed \cite{oscill. exper.}. There is 
agreement that  oscillation is possible when different mass
neutrino states add coherently \cite{Kayser} in their production
process. Such a combination of eigenmass neutrino states was defined
many years ago by Maki, Nakagawa and Sakata \cite{MNS} using
the previous idea of neutrino mixing invented by
Pontecorvo\cite{Pontecorvo}. Now, there is agreement
concerning the definition of so-called neutrino flavour states for
relativistic neutrinos.  The state of three massive relativistic
neutrinos produced in the process involving a charged flavour $ \alpha $
lepton is given by \cite{theory of oscillation}
\begin{eqnarray}
\label{MNSP} |\nu_{\alpha}\rangle =\sum_{i=1}^{3} U_{\alpha i}^{*}
|\nu_{i}\rangle,
\end{eqnarray}
where $U$ is the unitary mixing matrix known as the
Maki-Nakagawa-Sakata-Pontecorvo matrix.  Such a state, which
describes neutrinos with negative helicity, is understandable.
Neutrinos are produced in charged lepton processes, which are
described by the chiral left - handed interaction; the only one
existing in the Standard Model (SM). Coupling of the charged
$l_{\alpha}$ lepton with a massive $\nu_{i}$ neutrino and W boson is
given by:
\begin{eqnarray}
\label{elementary coupling} L_{CC} = \frac{e}{2 \sqrt{2} \ sin
\theta_{W}}\sum_{\alpha, i} \overline{\nu}_{i} \gamma^{\mu}
(1-\gamma_{5}) U_{\alpha i}^{*} l_{\alpha}  W_{\mu}^{+}  +  h.c.
\end{eqnarray}
Such an interaction produces a relativistic neutrino with negative
helicity (antineutrino with positive helicity), and its propagation
in a vacuum or in matter does not reverse spin direction (neutral
current interaction is also left-handed). In such circumstances,
it is natural to describe neutrinos by a pure quantum mechanical
state Eq.(\ref{MNSP}). So far, only relativistic neutrinos are
detected (the detection energy threshold is practically larger then
100 keV \cite {en. threshold}), and the pure production state
Eq.(\ref{MNSP}) works well. However, there are several reasons to
pose the following question about general neutrino states: What is the neutrino
state if it is not necessarily relativistic, and/or its
interaction is more general than that given by Eq.(\ref{elementary
coupling})? The answer for such a question is important as attempts
are ongoing to define different neutrino states in a non-relativistic
region (see e.g. \cite {Fock space} ). Besides, there is a chance that
future more precise neutrino oscillation experiments will open a window for New Physics (NP) 
and the knowledge of the initial and final neutrino states could be important.
The attempts to define general neutrino states have been discussed many times
in the literature (see e.g.\cite{Giunti I}). The neutrino production and detection states
that depend on the specific physical process were considered in
e.g. \cite {Grossman}. To our knowledge, the most general
oscillation and detection neutrino states were studied by Giunti
\cite {Giunti II} (see also \cite{others}). However, even in this
work, only pure states in the relativistic case, without the spin effect
of particles which produces neutrinos, are finally discussed.

The aim of our study is to construct a general neutrino density matrix using
the helicity - mass basis using the amplitude for the production
process in the CM system or rest system of a decaying object (e.g.
muon in the neutrino factory or nucleus in the beta beam
experiments) using a full wave packet description of particles.
In order to find how physics beyond the SM modify neutrino
production states, we consider the general neutrino interaction. Next,
we transform the density matrix to the frame of the detector - 
the laboratory frame. Such a general Lorentz transformation causes
Wigner rotation for spin states \cite{Wigner}. We show that,
even if such a Wigner rotation can be large, its effect on
neutrino helicity states is small and without practical meaning.
Then we have the general density matrix describing a neutrino beam,
which oscillates as it passes between the production
site and the detector. In this way, we will be able, although it is not
subject of this paper, to consider more precisely any possible
influence of NP not only on oscillation phenomena as 
usual (\cite{Oscillation}), but also on the production and,
in a similar way, detection of neutrino. In such a case, all quantum
mechanical properties of neutrino oscillation phenomena (such as
 the problem of factorization, oscillation and coherence
length) could be considered in the more precise way.

We ask questions about when the neutrino state is pure and when it must
be consider as mixed, how this property depends on neutrino energy,
and the mechanism of the production process. We show that, for
non-relativistic neutrinos, their state is always mixed. This is a
very natural result, for neutrinos with an energy comparable to their
masses, even a left-handed interaction (Eq.(\ref {elementary
coupling})) produces neutrinos with both helicities. Such a case
must be described by a mixed state. But, as we will see, the appearance
of the subdominant neutrino helicity states is not the only reason 
for the mixture of states. If the NP introduces different mixing
matrices for different kinds of interactions (for
vector and scalar couplings, as examples), quantum mixing of states also appears
. Because of that, any attempt to treat a flavour neutrino as
one object in Fock space (see e.g. \cite {Fock space}), even if
mathematically consistent, creates a problem with the description of the real
production process. Neutrino states are always mixed if the production
mechanism is more complicated in comparison with the SM
left-handed interaction Eq.(\ref {elementary coupling}). The
neutrino density matrix also depends on the other properties of the
production process; for example, the scattering angle,
polarizations of other accompanying particles, and their momentum
uncertainties as well as on the neutrinos themselves, the mass of the
lightest neutrino, or their mass hierarchy.

In this paper, we discuss only the first stage of the oscillation
process - the preparation of neutrino states - leaving the problem of
oscillation, together with the interaction inside the medium and the
problem of their detection, for a future paper.  In the next Chapter,
we describe the neutrino density matrix for the selected
production process. As the base process, we take the neutrino
production where charged flavour leptons $\alpha$  scatter on
nuclei A ($l_{\alpha} + A \rightarrow \nu_{i} + B$). Amplitudes
are calculated for the general interaction and the normalized ($Tr(\rho)
=1$) density matrix in the CM system is found. Then, the problem of the
Lorentz transformation and Wigner rotation is discussed. The full
analytical and numerical properties of the neutrino density matrix are
analyzed. To distinguish between mixed and pure states, we
calculate $Tr(\rho^{2})$ as  a function of neutrino energy and the
parameters which describe the SM extensions. We also discuss how
fast the neutrino states become pure as the energy increases. We study
density matrices for two subsystems: the first which describes both
neutrinos’ helicity independent of their mass, and the second
describing the mixture of three neutrinos, independent of their spin
state. Finally, Chapter 3 gives our final comments and future
prospectives.

\section{Neutrino states in the production processes}
\label{NP} Let us assume that the beam of neutrinos with mass i,
$\nu_{i}$ is produced by charged leptons $(\alpha = e,\mu)$
which scatter on nuclei A.
\begin{eqnarray}
\label{1}
 l_{\alpha} + A \rightarrow \nu_{i} + B .
\end{eqnarray}
We do not measure separate neutrinos and their masses; in fact, we do not
directly measure neutrinos at all. Quantum neutrino states can be
deduced from the measured states of all other particles accompanying
the neutrino.  So, having information about particles A, B and,
especially, about charged lepton $l_{\alpha}$ in the process
Eq.(\ref{1}), gives us the ability to define neutrino states. To define the
spatial or momentum distribution for neutrinos, we have to assume
that all particles in the discussed process do not have a precisely 
determined momentum, so they must be defined by wave packets and
not by plane waves (as is usually the case in other scattering processes). In
order to discuss coherence effects in the detection process, we
have to define the momentum distribution of the mass-helicity density
matrix.  Such a density matrix can be constructed from the amplitude
for the process (\ref{1}) whose
interaction mechanism is under investigation.

We assume that our interaction is described by an effective model
Lagrangian where both the vector and scalar couplings with the left and
right chirality interaction are present. This sort of interaction
appears, for example, in the Left - Right symmetric model (see e.g.
\cite {LR}). We assume that:
\begin{eqnarray} \label{charge current}
{\cal L}_{CC} = &-& \frac{e}{2\sqrt{2} sin\theta_W} \biggl\{
\sum_{\alpha ,i} \bar{\nu_i}\;[\;\gamma^{\mu} (1 - \gamma_5)
\epsilon_{L}^{c} U_{\alpha i}^{L \ast} + \gamma^{\mu} (1 +
\gamma_5) \epsilon_{R}^{c} U_{\alpha i}^{R \ast} ]\;l_{\alpha}\;
W^{+}_{\mu} \nonumber
\\
&+&  \sum_{\alpha ,i} \bar{\nu_i}\;[\; (1 - \gamma_5) \eta_{L}
V_{\alpha i}^{L \ast} + (1 + \gamma_5) \eta_{R} V_{\alpha i}^{R
\ast} ]\;l_{\alpha}\; H^{+}  \nonumber
\\
&+& \sum_{u ,d} \bar{u}\;[\;\gamma^{\mu} (1 - \gamma_5)
\epsilon_{L}^{q} U_{ud}^{\ast} + \gamma^{\mu} (1 + \gamma_5)
\epsilon_{R}^{q} U_{ud}^{\ast} ]\;d\; W^{+}_{\mu} \nonumber
\\
&+&  \sum_{u ,d} \bar{u}\;[\;(1 - \gamma_5) \tau_{L}  W_{ud}^{L
\ast} + (1 + \gamma_5) \tau_{R} W_{ud}^{R \ast} ]\;d\;  H^{+}
\biggr\} + h.c,
\end{eqnarray}
where $\varepsilon_{L,R}$, $\eta_{L,R}$ and $\tau_{L,R}$ are
parameters that are taken to define the scale and differ slightly from
their $\nu SM$ values.

In the neutrino sector, we introduce the mixing matrices separately
for chiral-left and chiral-right parts as well as for vector and
scalar neutrino interactions. As a consequence, phenomena such as
flavour neutrino states, which were explicitly defined, do not exist any
more, even for relativistic neutrinos. Mixing in the left-handed
vector interaction $(U^{L})$ and left-handed scalar one $(V^{L})$
can be different and an unambiguous definition of flavour neutrino
state is impossible. We see that the determination of the flavour neutrino
state given by the Pontecorvo-Maki-Nakagawa-Sakata matrix is purely
accidental and valid only in the SM and relativistic neutrino
interaction. In such circumstances, we cannot consider  flavour
neutrinos as elementary quanta of a field theory.
\\
\indent  In the lowest order, two  Feynman diagrams describe the
neutrino production process Eq.(\ref{1})
\\
\begin{figure}[h]
\begin{center} \begin{picture}(400,150)(-10,0)

 \ArrowLine(10,140)(50,100)
    \ArrowLine(50,100)(90,140)
    \Photon(50,100)(50,50){2}{3}\Text(40,80)[]{$q^{2}$}\Text(80,80)[]{$W^{+}, H^{+}$}
    \DashLine(50,100)(50,50){3}
    \ArrowLine(50,74)(50,78)
    \ArrowLine(10,10)(50,50)
    \ArrowLine(50,50)(90,10)
    \Text(0,140)[]{$l_{\alpha}$}
    \Text(0,0)[]{$A$}
    \Text(100,140)[]{$\nu_{i}$}
    \Text(100,0)[]{$B$}
    \Text(150,-30)[]{}

    \end{picture}
    \end{center}
\caption{\small The charged boson(s) $W^{+}$ and charged Higgs
particle(s) $H^{+}$ exchange diagrams, which describes the neutrino
production process.}
\end{figure}
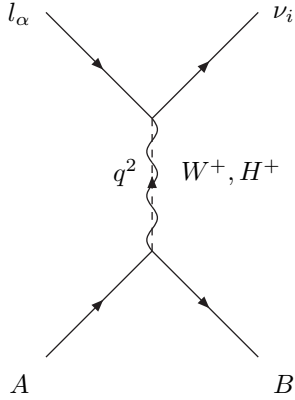

To define the neutrino state we have to calculate the amplitude
for the process Eq. (\ref{1}). Three measured particles
$l_{\alpha}$, A and B are in the states given by the wave packets
$(a = l, A,B)$
\begin{eqnarray}
\label{state} |\psi_{a}, \lambda_{a}\rangle = \int d^{3}p_{a}
\Psi_{a}(\vec{p}_{a}, \vec{p}_{a0},\sigma_{ap}) |\vec{p}_{a},
\lambda_{a}\rangle ,
\end{eqnarray}
where $|\vec{p}_{a}, \lambda_{a}\rangle$ is the particle state
with definite momentum $\vec{p}_{a}$ and helicity $\lambda_{a}$,
and $\Psi_{a}(\vec{p}_{a}, \vec{p}_{a0},\sigma_{ap})$ are the
momentum distribution wave functions with a central momentum
$\vec{p}_{a}$ and momentum uncertainty $\sigma_{ap}$. For
A narrow momentum distribution ($\sigma_{ap} \ll |\vec{p}_{a}|$), the
physical results are independent of the specific shape of these
functions, and we take them in the Gaussian form:
\begin{eqnarray}
\label{state} \Psi_{a}(\vec{p}_{a}, \vec{p}_{a0},\sigma_{ap})=
\frac{1}{(2 \pi \sigma_{ap}^{2})^{\frac{3}{4}}} e^{ -
\frac{(\vec{p}_{a} - \vec{p}_{a0})^{2}}{4 \sigma_{ap}^{2}}}.
\end{eqnarray}
With such wave packet states, we calculate the amplitude for the
process (\ref{1}). After integration over the particles’ momenta we
obtain:
\begin{eqnarray}
\label{2}\langle f | S | i \rangle = -i \frac{1}{[(2 \pi
\sigma_{lx}^{2})(2 \pi \sigma_{Ax}^{2})(2 \pi
\sigma_{Bx}^{2})]^{\frac{3}{4}}} \int d^4 x e^{- i (E-E_{R})t}
e^{i(\vec{p}-\vec{p_R})\vec{x}}e^{- \frac{\vec{x}^2-2\vec{v}_R
\vec{x}t+\Sigma_R t^2}{\sigma_{Rx}^2}}M(\vec{p}, \lambda, ...),
\end{eqnarray}
where the average values of neutrino energy and momentum are
determined from the appropriate conservation rules (we use the
same notation as in Ref (\cite {Giunti II})):
$$E_{R} = E({\vec{p}}_{l0}) + E({\vec{p}}_{A0}) -
E({\vec{p}}_{B0}^{\ i}),$$ and
$${\vec{p}}_{R} = {\vec{p}}_{l0} + {\vec{p}}_{A0} -
{\vec{p}}_{B0}^{\ i}.$$ For the precise energy and momentum
determination we have to remember that, for a given total energy in
the process (\ref{1}), the final particle momenta and energies
depend on the neutrino masses. The neutrino spatial uncertainty
$\sigma_{Rx}$ and average velocity $\vec{v}_{R}$ are obtained from the
accompanying particles uncertainties $\sigma_{ax} = \frac{1}{2
\sigma_{ap}}$ and their velocities $\vec{v}_{a} =
(\frac{dE}{d\vec{p}})_{\vec{p}= \vec{p}_{a}}$:
$$\frac{1}{\sigma^{2}_{Rx}} = \frac{1}{\sigma^{2}_{lx}} + \frac{1}
{\sigma^{2}_{Ax}} + \frac{1}{\sigma^{2}_{1Bx}},$$
$${\vec{v}}_{R} = \sigma^{2}_{Rx}\left( \frac{{\vec{v}}_{l0}}{\sigma^{2}_{lx}}
 + \frac{{\vec{v}}_{A0}}{\sigma^{2}_{Ax}}
+ \frac{{\vec{v}}_{B0}}{\sigma^{2}_{Bx}}\right).$$ And, finally,
$${\Sigma}_{R} = \sigma^{2}_{Rx}\left( \frac{{\vec{v}}^{2}_{l0}}{\sigma^{2}_{lx}}
+ \frac{{\vec{v}}^{2}_{A0}}{\sigma^{2}_{Ax}} +
\frac{{\vec{v}}^{2}_{B0}}{\sigma^{2}_{Bx}} \right).$$ The
amplitudes $M(\vec{p},\lambda,...) \equiv M(\vec{p},
\lambda;\lambda_{l},\lambda_{A}, \lambda_{B})$ are  the full
helicity amplitudes for the production process  Eq.(\ref{1}), which is
calculated in the CM frame for average values of all particle
momenta except for the neutrino. Using the saddle-point approximation,
the integrations over space and time can be done, giving the final
amplitude $A^{\alpha}_{i}(\vec{p},\lambda, \lambda_{B};
\lambda_{l}, \lambda_{A})$ for the production of the $\nu_{i}$
neutrino with mass $m_{i}$, momentum $\vec{p}$, and helicity
$\lambda$ in the process initiated by flavour $\alpha$ lepton:
\begin{eqnarray}
\label{Amplitude}A^{\alpha}_{i}(\vec{p},\lambda, \lambda_{B};
\lambda_{l}, \lambda_{A}) \ =  \ n_{\alpha i} \  f_{\alpha
i}(\vec{p},E) \  M^{\alpha}_{i}(\vec{p},
\lambda;\lambda_{l},\lambda_{A}, \lambda_{B}),
\end{eqnarray}
where $$ n_{\alpha i} = \frac{2^{\frac{9}{4}} ( \sigma_{lp} \
\sigma_{Ap} \
\sigma_{Bp})^{\frac{3}{2}}}{\pi^{\frac{1}{4}}\sigma_{Rp}^4\sqrt{\Sigma_R-v_R^2}},$$
 and
$$f_{\alpha i}(\vec{p},E) = e^{- \frac{({\vec{p}} - {\vec{p}}_{R})^{2}}{4{\sigma}^{2}_{Rp}} -
 \frac{({\vec{v}}_{R}({\vec{p}} - {\vec{p}}_{R}) - E + E_{R})^{2}}
 {\sigma^{2}_{Rp}\left({\Sigma}_{R} - {\vec{v}}^{2}_{R}\right)}},$$
where we have openly indicated that both factors depend on the
neutrino mass. Once more, momentum uncertainty $\sigma_{Rp} =
\frac{1}{2 \sigma_{Rx}}$ appears in the above formulas. The second
function $f_{\alpha i}$ describes the momentum distribution for the
produced neutrinos.

The amplitude Eq.(\ref {Amplitude}) describes all properties of the
produced neutrinos. Their quantum mechanical state is determined
by the density matrix defined as:
\begin{eqnarray}
\label{Density Matrix} \varrho^{\alpha}(\vec{p},\ i, \ \lambda ; \
\vec{p}^{\ '},i^{'},\lambda^{'}) =\frac{1}{N_{\alpha}}
\sum_{\lambda_{l},\lambda_{A},\lambda_{B}}
A^{\alpha}_{i}(\vec{p},\lambda, \lambda_{B}; \lambda_{l},
\lambda_{A})A^{\alpha *}_{i^{'}}(\vec{p}^{\ '},\lambda^{'},
\lambda_{B}; \lambda_{l}, \lambda_{A}),
\end{eqnarray}
where we assume that, in the production process (\ref {1}), the initial
particles are not polarized and the final polarization of the B nuclei
is not measured. For the normalization factor $N_{\alpha}$, given by
\begin{eqnarray}
N_{\alpha} = \sum_{\lambda = \pm 1} \sum_{i = 1}^{3}
\sum_{\lambda_{l},\lambda_{A},\lambda_{B}}\int d^{3}p
|A^{\alpha}_{i}(\vec{p},\lambda, \lambda_{B}; \lambda_{l},
\lambda_{A})|^{2},
\end{eqnarray}
the density matrix is properly normalized:
\begin{eqnarray}
Tr(\varrho^{\alpha})= \sum_{\lambda = \pm 1} \sum_{i = 1}^{3} \int
d^{3}p \ \varrho^{\alpha} (\vec{p},\lambda,i;\vec{p},\lambda,i) =
1.
\end{eqnarray}
Eq. (\ref{Density Matrix}) describes the neutrino state in the CM
frame. Such a frame is equivalent to the rest frame of a decaying muon
in the neutrino factory case or the decaying nuclei in the beta beam
experiments. Usually, neutrinos are not produced in such a frame.
They are produced in the laboratory frame (LAB), where decaying
muons or nuclei are moving very fast. The density matrix in such a frame
is interesting for us. To reach the LAB frame, we have to make a
Lorentz transformation. As helicity is not a full Lorentz scalar, we
can expect that the density matrix in the LAB will be rotated
(helicity Wigner rotation). Each neutrino state
$|\vec{p},\lambda,i\rangle$ with mass $m_{i}$, momentum $\vec{p}$
and helicity $\lambda$ after a general Lorentz transformation
$\Lambda$, does not necessary point along the neutrino momentum,
but transforms as:
\begin{eqnarray}
\label{2}U[\Lambda] |\vec{p},s,\lambda\rangle = \sum_{\sigma}
\textbf{D}^{s}_{\sigma\,\lambda} \Big(
R^{T}(\varphi,\overline{\theta},-\varphi)R_{W}
R(\varphi,\theta,-\varphi) \Big)|\Lambda \vec{p},s,\sigma\rangle,
\end{eqnarray}
where angles $\varphi$ and $\theta$ $(\varphi$ and
$\overline{\theta})$ are the spherical angles of momenta before
$(\vec{p})$ and after $(\Lambda \vec{p})$  which carry the Lorentz
transformation. $R_{W}$ is the normal Wigner rotation:
\begin{eqnarray}
\label{RW}R_{W} = L^{-1}(\Lambda\vec{p}) \Lambda L(\vec{p}).
\end{eqnarray}
Hence, each element of the density matrix transforms as:
\begin{eqnarray}
\label{LT}\varrho^{\alpha}(\vec{p},\ i, \ \lambda ; \ \vec{p}^{\
'},i^{'},\lambda^{'})\rightarrow \varrho^{\alpha}(\Lambda
\vec{p},\ i, \ \eta ; \ \Lambda \vec{p}^{\ '},i^{'},\eta^{'}) = \\
\nonumber
\\ \nonumber  \, \, \, \,  \sum_{\lambda \lambda^{'}} \textbf{D}^{1/2}_{\eta
\,\lambda}(R_{W i}^{h}) \, \varrho^{\alpha}(\vec{p},\ i, \ \lambda
; \ \vec{p}^{\ '},i^{'},\lambda^{'}) \, \textbf{D}^{1/2
*}_{\eta^{'} \,\lambda^{'}}(R_{W i^{'}}^{h}), \nonumber
\end{eqnarray}
where $R_{W i}^{h}$  is the helicity Wigner rotation:
$$R_{W i}^{h}=R^{T}(\varphi,\overline{\theta},-\varphi)R_{W}
R(\varphi,\theta,-\varphi),$$ and depends on neutrino mass as
the neutrino momentum depends on it. The matrix
$\textbf{D}^{1/2}_{\eta \,\lambda}(R_{W i}^{h})$ can be found. If
we parameterize:
\begin{eqnarray}
\label{21}   \textbf{D}^{1/2}(R_{W i}^{h})=\left(%
\begin{array}{cc}
  a & b \\
  -b^{*} & a^{*} \\
\end{array}%
\right),
\end{eqnarray}
then $$b=|b|e^{-i \varphi},$$ where
$$|b|=-\frac{S( \tilde{p} +C \tilde{E} \cos (\theta ))-C \tilde{E}
\sin (\theta )} {\sqrt{2 [(\tilde{E} +\tilde{p}  \cos (\theta ))
(\tilde{E} +C \tilde{p} +\tilde{p}  \cos (\theta ) + C \tilde{E}
\cos (\theta )+S \tilde{E} \sin (\theta ))]}},$$
$$a = \sqrt{1 - |b|^{2}},$$ and
$$\tilde{p} = p \beta  \gamma, \, \, \, \,  \tilde{E} = m_{i}+E \gamma,$$ with
$$C = \frac{\gamma  (E \beta +p \cos (\theta ))} {\sqrt{\gamma^2
(E \beta +p \cos (\theta ))^2 +p^2 \sin ^2(\theta )}}, \, \, \, \,
\, S=\sqrt{1-C^2}.$$ Here $m_{i}, E$ and $p$ are the neutrino mass,
energy and momentum, $\gamma$ and $\beta$ are parameters of
the Lorentz transformation in the "z" direction, and $\theta,\varphi$ are
spherical angles of the neutrino momentum.
\begin{figure}[h]
\centering
\includegraphics[scale=0.4]{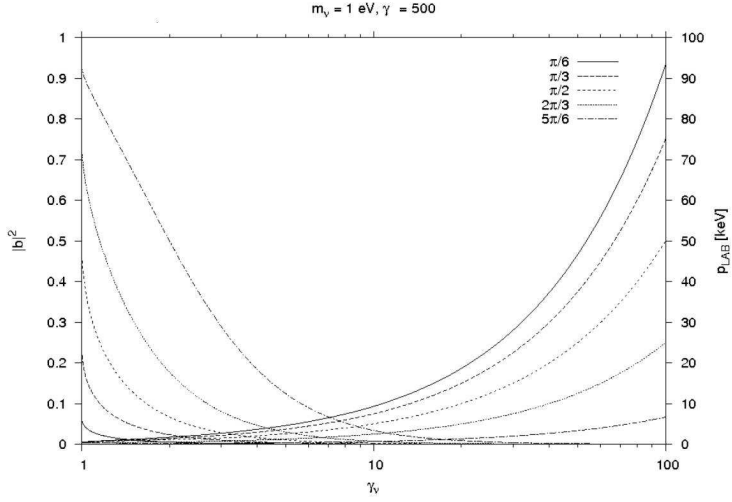}
\caption{{\small  On the left - the off-diagonal element $|b|$ of
the helicity Wigner rotation matrix as a function of the neutrino's
$\gamma_{\nu}= \frac{E}{\, m_{i}}$ factor for five different
neutrino scattering angle $\theta =
\frac{\pi}{6},\frac{\pi}{3},\frac{\pi}{2},\frac{2 \pi}{3}$ and
$\frac{5 \pi}{6}$ in CM. The Lorentz transformation is described by
the $\gamma = 500$ factor. On the right side, the neutrino momentum
in LAB system for the same scattering angle in the CM frame is shown.}}
\label{fig:2}
\end{figure}
It is worth noticing that, for normal (not helicity) Wigner
rotation $R_{W}$,  $C\rightarrow cos\theta$ and $S\rightarrow
sin\theta$. The off-diagonal element of the $D^{1/2}$ matrix Eq.(\ref
{21}) b decides how big the effect of the Wigner rotation is.
 In Fig. 2, on the left side, we have plotted
the $|b|$ element as a function of neutrino $\gamma$ factor
($\gamma_{\nu} =\frac{E}{m_{i}}$) for various scattering angles
$\theta_{CM}$ for the Lorentz transformation along the "z" axis with
$\gamma = 500$. We see that the effect is large but only for very
small neutrino energies ($E\approx m_{i}$) and large scattering
angles.  Unfortunately, as we see from right side of the picture,
such neutrinos also have small energies in LAB frame and are not
measured. The effect of the helicity flip for experimentally measurable
neutrinos with a enough large energy ($E_{LAB} \gtrsim 100 $ MeV) is
completely negligible. We see that, for practical purposes, we can
safely take it that Lorentz transformation does not change the
helicity structure of the density matrix and
\begin{eqnarray}
\label{DM in LAB}\varrho^{LAB}(\Lambda \vec{p},\ i, \ \lambda ; \
\Lambda \vec{p}^{\ '},i^{'},\lambda^{'}) =
 \, \, \, \,   \varrho^{CM}(\vec{p},\ i, \ \lambda
; \ \vec{p}^{\ '},i^{'},\lambda^{'}).
\end{eqnarray}
We can now check the quantum structure of the neutrino state
described by Eq.(\ref{Density Matrix}). First, we would like to
answer how good the MNSP assumption (Eq.(\ref {MNSP})) is. In order
to do that, we calculate $Tr[(\varrho^{\alpha})^{2}]$:
\begin{eqnarray}
\label{DM in LAB}Tr[(\varrho^{\alpha})^{2}]= \sum_{\lambda\, i \,
\lambda^{'} i^{'}}\int d^{3}p \  \,  d^{3}p^{\, '}
  \varrho^{\alpha}(\vec{p},\ i, \ \lambda
; \ \vec{p}^{\ '},i^{'},\lambda^{'}) \varrho^{\alpha}(\vec{p}^{\
'}, i^{\ '},  \lambda^{'} ; \ \vec{p},\ i,\ \lambda), \,
\end{eqnarray}
and check how it differs from unity.

To be more precise and to not complicate our consideration with the
structures of particles A and B, which are not important for our
purpose, we take the u and d quarks with
masses $m_{u} = 2.25 \  MeV$ and $m_{d} = 5.0 \  MeV$ as particles A and B, respectively. For lepton
$\alpha$ we take an electron with $m_{e} = 0.511 \ MeV$. All
Particles’ momenta uncertainties are equal and $\sigma_{ep} =
\sigma_{up} = \sigma_{dp} = 10 \ eV$. The results do not depend
crucially on the values of these uncertainties.
\begin{figure}[h]
\centering
\includegraphics[scale=0.40]{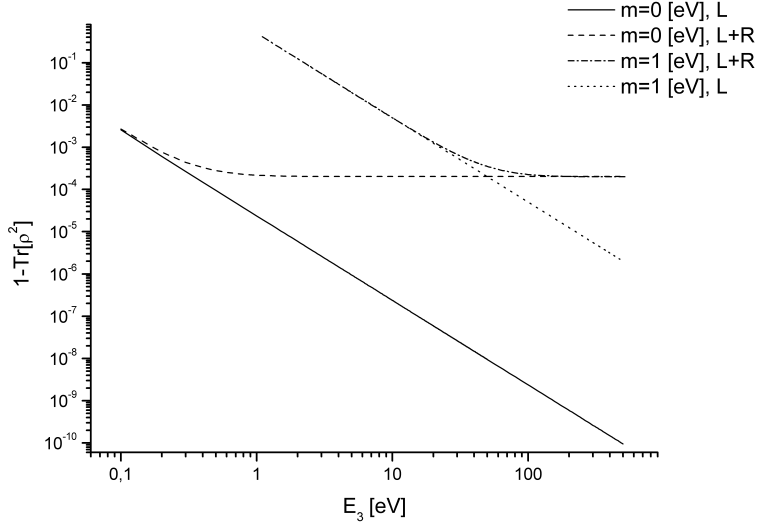}
\caption{\small  $1-Tr[\rho^2]$ as a function of the neutrino kinetic
energy for the two lightest neutrino mass $m_{1} = 0 \ eV$ and $m_{1}
= 1 \ eV$ with only Left ($\varepsilon_L=1,\varepsilon_R=0.)$ and the
Left- and Right-currents ($\varepsilon_L=1,\varepsilon_R=0.01$).}
\label{fig:3}
\end{figure}
In Fig. 3, we plot $1 - Tr[(\varrho^{\alpha})^{2}]$ as a function of
the kinetic neutrino energy for two values of the lightest
neutrino mass, $m_{1} = 0 \ eV$ and $m_{1} = 1 \ eV$ in two cases,
when only Left-Handed chiral interactions are present (only
$\varepsilon_{L} \neq 0$), and when Left- and Right-current exist
($\varepsilon_{L} \neq 0$ and $\varepsilon_{R} \neq 0$). The masses of
the heavier neutrinos and the vector Left-Handed mixing ($U_{L}$)
are taken from the oscillation data \cite{Oscillation data}. We
assume a normal mass hierarchy with $m^{2}_{\odot}$ = $8.0 \times
10^{-5} eV^{2}$, $m^{2}_{atm}$ = $2.5 \times 10^{-3} eV^{2}$ and
mixing angles, $sin^{2}\theta_{12} = 0.31, \, sin^{2}\theta_{23} =
0.44, \, sin^{2}\theta_{13} = 0.009$ and $\delta_{CP} =0$ (central
values are taken). The mixing in the Right-Handed vector ($U_{R}$)
and scalar interactions ($V_{L,R}$) we take to be the same as in the L-H
case ($V_{L} = V_{R} = U_{R} = U_{L}$). As we will see, by taking
the same mixing matrices, we ignore the QM mixing caused by
different mass states, and assume possible mixing effects are caused only
by different helicity states. If the production process is
described by the SM interaction (only L-H vector interaction), the
state is strongly mixed for very small neutrino energies (continuous
line and dotted line). The effect of the mixture depends on the
lightest neutrino mass (as we see from the Fig. 3) and depends
also slightly on the mass hierarchy  and chosen particle momentum
uncertainties (not shown).

If the neutrinos’ energies increase, their state becomes pure. For
relativistic neutrinos where masses and momentum uncertainties can
be neglected, the density matrix takes the form:
\begin{eqnarray}
\label{Relativistic Case ++} \varrho^{\alpha}(+1,i;+1,k) =
\frac{1}{N_{\alpha}}\big(A_{\varepsilon_{R}^{2}} \,
\varepsilon_{R}^{2} \, U_{\alpha i}^{R *} U_{\alpha k}^{R} +
A_{\eta_{L}^{2}} \, \eta_{L}^{2} \, V_{\alpha i}^{L *} V_{\alpha
k}^{L} + A_{\varepsilon_{R}\eta_{L}} \,  \varepsilon_{R} \eta_{L}
\,( U_{\alpha i}^{R *} V_{\alpha k}^{L } + V_{\alpha i}^{L *}
U_{\alpha k}^{R})\big),
\end{eqnarray}
\begin{eqnarray}
\label{Relativistic Case+-} \varrho^{\alpha}(+1,i;-1,k)
=\varrho^{\alpha}(-1,k;+1,i)^{*} = \\
\frac{1}{N_{\alpha}}\big(A_{\varepsilon_{L} \varepsilon_{R}} \,
\varepsilon_{R} \varepsilon_{L} \, U^{R *}_{\alpha i}
U^{L}_{\alpha k} + A_{\varepsilon_{R} \eta_{R}} \, \varepsilon_{R}
\eta_{R} \, U^{R *}_{\alpha i} V^{R}_{\alpha k} +
A_{\varepsilon_{L} \eta_{L}} \, \varepsilon_{L} \eta_{L} \,  V^{L
*}_{\alpha i} U^{L}_{\alpha k}\big), \nonumber
\end{eqnarray}
\begin{eqnarray}
\label{Relativistic Case--} \varrho^{\alpha}(-1,i;-1,k) =
\frac{1}{N_{\alpha}}\big(A_{\varepsilon_{L}^{2}} \,
\varepsilon_{L}^{2} \, U_{\alpha i}^{L *} U_{\alpha k}^{L} +
A_{\eta_{R}^{2}} \, \eta_{R}^{2} \, V_{\alpha i}^{R *} V_{\alpha
k}^{R} + A_{\varepsilon_{L}\eta_{R}} \, \varepsilon_{L} \eta_{R}
\,( U_{\alpha i}^{L *} V_{\alpha k}^{R} + V_{\alpha i}^{R *}
U_{\alpha k}^{L})\big),
\end{eqnarray}
and
\begin{eqnarray}
\label{Relativistic Case--} \ \ N_{\alpha} =
A_{\varepsilon_{L}^{2}} \, \varepsilon_{L}^{2} \, +
A_{\eta_{R}^{2}} \, \eta_{R}^{2} \,  + A_{\varepsilon_{R}^{2}} \,
\varepsilon_{R}^{2} \,  + A_{\eta_{L}^{2}} \, \eta_{L}^{2} \, +
\\ A_{\varepsilon_{R}\eta_{L}} \,  \varepsilon_{R} \eta_{L} \,(
V^{L} U^{R \dag } +U^{R} V^{L \dag} )_{\alpha\alpha}
+A_{\varepsilon_{L}\eta_{R}} \, \varepsilon_{L} \eta_{R} \,( V^{R}
U^{L\dag} + U^{L} V^{R\dagger})_{\alpha\alpha}, \nonumber
\end{eqnarray}
 where the A factors (not given here) depend on the neutrino energy and
scattering angles. We see that, in the SM where only
$\varepsilon_{L} \neq0$:
\begin{eqnarray}
\label{Relativistic Case--} \varrho^{\alpha}(-1,i;-1,k) =
 U_{\alpha i}^{L
*} U_{\alpha k}^{L},
\end{eqnarray}
and all other elements of density matrix are equal zero. Such a
density matrix describes pure states equivalent to the MNSP mixture
Eq. (\ref{MNSP}). \textbf{Relativistic neutrinos in the SM model
are produced in the pure states described by the MNSP mixture}. The
situation changes if any NP participates in the neutrino
production process. Then, the states are not pure - even for
relativistic neutrinos. A departure from the pure state depends on
two properties of the system: the helicity mixing and the
mass mixing. The first is obvious - neutrinos in two helicity
states are produced which, in a natural way, gives the mixed state. The
second effect is more mysterious and appears only if the mixing
matrices are not the same. In order to investigate both effects, we
can construct separately two density matrices for two subsystems -
the first for the helicity subsystem and the second for the mass. For
the helicity subsystem, the density matrix is given by:
\begin{eqnarray}
\label{Helicity DM}\varrho^{\alpha,spin}_{\lambda,\eta} =
\sum_{i=1}^{3} \varrho^{\alpha}(\lambda,i;\eta,i),
\end{eqnarray}
and, for the mass subsystem:
\begin{eqnarray}
\label{Mass DM}\varrho^{\alpha,mass}_{i,k} = \sum_{\lambda=\pm 1}
\varrho^{\alpha}(\lambda,i;\lambda,k).
\end{eqnarray}
With reference to the helicity subsystem, $1 -
Tr[(\varrho^{\alpha,spin})^{2}]$ is different then zero even if
all mixing matrices are equal, and it is a second order effect in the NP
parameters: $$1-Tr[(\varrho^{\alpha,spin})^{2}]\approx
O(\varepsilon_{R}^{2},\eta_{R}^{2},\eta_{L}^{2}).$$ In Fig. 3, we
see (dashed line $(m_{1}=0 \, eV)$ and dashed - doted line $(m_{1}
= 1 \, eV)$ that above the non-relativistic region, where even the L-H
vector current produces neutrinos with h=+1/2,  the state mixture
is almost constant and of the order of $\varepsilon_{R}^{2}  \approx$
0.0001. Generally, the mixture of the neutrino states is, in a good
approximation, proportional to the probability squared that
neutrinos with negative helicity are produced.

\begin{eqnarray}
\label{Relation mixture probability} Tr[(\varrho^{\alpha})^{2}]
\approx (Tr[ \varrho^{\alpha} \widehat{P}_{-}])^{2},
\end{eqnarray}
where the $\widehat{P}_{-}$ is the projection operator on the
negative helicity state, $(\widehat{P}_{-})_{\lambda, \, i \, ;\,
\eta,\, k}=\delta_{\lambda,-1} \delta_{\eta, -1} \delta_{i,k}$.
\begin{figure}[h]
\centering
\includegraphics[scale=0.2]{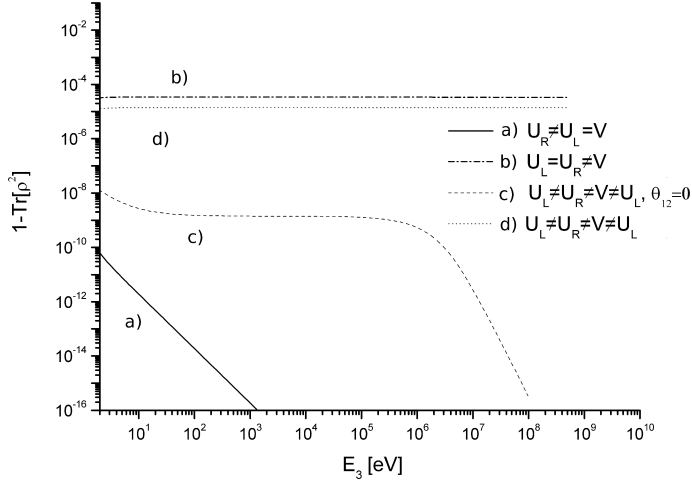}
\caption{{\small 1-$Tr[(\varrho^{\alpha,mass})^{2}]$ as a function
of the kinetic neutrino energy for various mixing matrices in the
vector and scalar Lagrangian.  For the parameter which defines New
Physics, we take $\varepsilon_{R} = 0.01, \eta_{L} = \eta_{R} =
0.02$.  All matrices are parameterized in the same way as the
standard $U_{L}$ matrix. \,  a) $sin^{2}\theta_{12} = 0.87,
sin^{2}\theta_{23} = 0.09, sin^{2}\theta_{13} = 0.23$,
$\delta_{CP} = 0.5$ for $U_R$ and $V_{L} = V_{R} = U_{L}$; \, b)
$V_{L} = V_{R}$ as $U_{R}$ in the case a) and $U_{R} = U_{L}$; \,
c) $V_{L} =V_{R}$ the same as in b) and $sin^{2}\theta_{12} = 0,
sin^{2}\theta_{23} = 0.09, sin^{2}\theta_{13} = 0.71$,
$\delta_{CP} = 0.8$ for $U_R$; \, d) $V_{L}= V_{R}$ as in the case
b) and $sin^{2}\theta_{12} = 0.01, sin^{2}\theta_{23} = 0.09,
sin^{2}\theta_{13} = 0.71$, $\delta_{CP} = 0.8$ for $U_R$.
 }} \label{fig:4}
\end{figure}
In Fig. 4, we depicted the effect of neutrino state mixtures caused
by the mass mixing. First, for relativistic neutrinos, there
is no mixture if all mixing matrices are the same: $$
Tr[(\varrho^{\alpha,mass})^2] = 1,$$ if $U_{R} = V_{R} = V_{L} =
U_{L}$. We see also that the size of the mixture depends on the shape
of the chosen mixing matrices. For some mixing matrices, the
mixture can be large and of the order of
$O(\varepsilon_{R}^{2},\eta_{R}^{2},\eta_{L}^{2})$.

Both sources of state mixtures, helicity mixing and mass
mixing, can have comparable effects. For example, the dotted-dashed
line in Fig. 4, where we take $U_{R}= U_{L}\neq V_{L} = V_{R}$,
indicates that the departure of $Tr[(\varrho^{\alpha, mass})^{2}]$
from unity is maximal. Here, we parameterize the V matrices in the same
way as the standard $U_{L}$ matrix and take $sin^{2}\theta_{12} =
0.87, sin^{2}\theta_{23} = 0.09, sin^{2}\theta_{13} = 0.23$ and
$\delta_{CP} = 0.5$. If, instead, we take $V_{L} = V_{R} = U_{L}
\neq U_{R}$ (the continuous line in Fig. 4), the mixture caused by
mixing between different states of neutrino masses is marginal. The
neutrino mixing in the scalar sector has a decisive role.

\section{Conclusions}
\label{CONCL} We have constructed the density matrix for neutrinos
created  in any  production process. Numerical calculations have
been  performed for the elementary reaction $e^{-} + u \rightarrow
\nu_{i} + d$. The density matrix has been calculated in the total
Center of Mass system. Beyond this, we have shown that the Lorentz
transformation to Laboratory system (the rest frame of a detector)
has negligible effects on the density matrix for experimentally interesting
neutrinos. For neutrinos with energies close
to their masses, their states must be described by a mixed density
matrix. In this region, the effect depends on the lightest neutrino
mass, the mass hierarchy, and the particles’ momenta uncertainties
given in wave packets. It was shown that, for relativistic
neutrinos, mixing of the states depends on the mechanism of the
production process. For the $ \nu SM $, where only a Left-Handed
vector current is in effect, neutrinos
are produced in a pure state with extreme accuracy. In this way, the assumption about
so-called flavour states as in the Maki-Nakagawa-Sakata-Pontecorvo
mixture of the  mass states, is justified. If NP
interactions also influence the neutrino production
process, the assumption about purity of neutrino production states
is not legitimate. Departure from pure states is caused by two
things: the possible production of neutrinos in two helicity
states caused, for example, by Right-Handed currents or by scalar
interactions, and by their mixing given by different mixing matrices
in the various sectors of the production mechanism. Separate
density matrices for the helicity subsystem and mass subsystem were
constructed. Both of these effects have been investigated numerically
and analytically in the case of relativistic and non-relativistic
neutrinos.

\textbf{Acknowledgments}: This work has been partially supported
by the Polish Ministry of Science under grant No.1 P03 B 049 26,
and by the European Community's Marie-Curie Research Training Network
under contract MRTN-CT-2006-035505 "Tools and Precision Calculations 
for Physics Discoveries at Colliders"


\begin{thebibliography}{99}
\bibitem{oscill. exper.}Fukuda, Y. et al. (Super-Kamiokande), Phys. Rev. Lett.
81 (1998) 1562-1567; Davis, R., Prog. Part. Nucl. Phys. 32 (1994)
13-32; Hosaka, J. et al. (Super-Kamkiokande), Phys. Rev. D73
(2006) 112001; Phys. Rev. Lett. 89 (2002) 011301.
\bibitem{Kayser}B. Kayser, Phys. Rev. D 24 (1981) 110.
\bibitem{MNS} Z. Maki, M. Nakagawa and S. Sakata, Prog.Theor. Phys.
28 (1962) 870.
\bibitem{Pontecorvo} B. Pontecorvo, J. Expetl. Theoret. 33 (1957) 549 and ibid.
34 (1958) 247.
\bibitem{theory of oscillation} S. Eliezer and A.R. Swift, \ Nucl. Phys.
B105 (1976) 45; \ H. Fritzsch and P. Minkowski,\  Phys.Lett. B62
(1976) 72; \ S.M. Bilenky and B. Pontecorvo,\ Sov.  J. Nucl.Phys.
24 (1976) 316; \ S.M. Bilenky and B. Pontecorvo,\  Nuovo Cim.
Lett. 17 (1976) 569.
\bibitem{en. threshold} see e.g. S. Bilenky, C. Giunti and W. Grimus, Prog.
Part. Nucl. Phys. 43 (1999) 1.
\bibitem{Fock space} M.Blasone, G.Vitiello, Annals Phys. 244
(1995) 283; \,  Blasone, P.A.Henning and G. Viliello, Phys.Lett.B
451 (1999) 140; \, K.Fujii, C.Habe and T. Yabuki, Phys.Rev. D59
(1999) 113003 and  Phys.Rev.D64 (2001) 013011; \, M.Binger and
C.R.Ji, Phys.Rev. D60 (1999) 056005;\, C.R.Ji and Y.Mishchenko,
Phys.Rev. D64 (2001) 076004; \ D.Boyanovsky and C.M.Ho, Phys.Rev
D69 (2004) 125012.
\bibitem{Giunti I}
Kiers, N. Weiss, Phys. Rev. D57 (1998) 3091;  C. Giunti, J.Phys.
G34 (2007) 93; C.Giunty, Eur. Phys. J. C39 (2005) 377.
\bibitem{Grossman} Y.Grossman, Phys.Lett. B359 (1995) 141.
\bibitem{Giunti II}  C. Giunti, JHEP 0211 (2002) 017.
\bibitem{others}M. Beuthe, Phys.Rept. 375 (2003) 105;\,  Ch. Cardall, Phys.Rev. D61 (2000)
073006; \, M. Zralek, Acta Phys. Polon. B29 (1998) 3925.
\bibitem{Wigner}E. P. Wigner, Ann. Math. 40, 149 (1939) 204.


\bibitem{Oscillation}
C. Giunti, C. W. Kim and U. W. Lee, Phys. Lett. B 274 (1992) 87;
 \, Y. Takeuchi, Y. Tazaki, S. Y. Tsai, T. Yamazaki, Prog. Theor.
Phys. 105 (2001) 471-482; \,   M. Nauenberg, Phys. Lett. B447
(1999) 23-30.

\bibitem{LR} J.C. Pati and A. Salam,Phys.Rev. D10 (1974) 277; R.N.
Mohapatra and J.C. Pati, Phys.Rev. D11 (1975) 2558; G. Sanjanovic,
Nucl.Phys Nucl. Phys. B153 (1979) 334; P.Duka, J. Gluza and M.
Zralek, Annals of phys. 280 (2000) 336.

\bibitem{Oscillation data}G.L. Fogli et all,
Prog.Part.Nucl.Phys. 57 (2006) 742-795.

\end{thebibliography}
\end{document}